\documentclass[aps,prd,showpacs,nopacs,nofootinbib,superscriptaddress,amsmath,amssymb]{revtex4}

\usepackage{graphicx}
\usepackage{graphics}
\usepackage{color}
\usepackage{epsfig}

\usepackage{graphicx}
\usepackage{amsmath}
\usepackage{amssymb}

\usepackage{amsfonts}
\usepackage{bm}
\usepackage{verbatim}

\usepackage{graphicx}
\usepackage{graphics}

\usepackage{color}

\usepackage{epsfig}

\begin{document}

\title{Analogies between the Black Hole Interior and the Type II Weyl Semimetals}

\author{M.A. Zubkov \footnote{On leave of absence from Institute for Theoretical and Experimental Physics, B. Cheremushkinskaya 25, Moscow, 117259, Russia}}
\email{zubkov@itep.ru}
\affiliation{Physics Department, Ariel University, Ariel 40700, Israel}

\date{28/11/2018}

\begin{abstract}
In the Painleve--Gullstrand (PG) reference frame, the description of elementary particles in the background of a black hole (BH) is similar to the description of non-relativistic matter falling toward the BH center. The velocity of the fall depends on the distance to the center, and it surpasses the speed of light inside the horizon.~Another analogy to non-relativistic physics appears in the description of the massless fermionic particle. Its Hamiltonian inside the BH, when written in the PG reference frame, is identical to the Hamiltonian of the electronic quasiparticles in type~II Weyl semimetals (WSII) that reside in the vicinity of a type~II Weyl point. When these materials are in the equilibrium state, the type II Weyl point becomes the crossing point of the two pieces of the Fermi surface called Fermi pockets. {It was previously stated} that there should be a Fermi surface inside a black hole in equilibrium. In real materials, type II Weyl points come in pairs, and the descriptions of the quasiparticles in their vicinities are, to a certain extent, inverse. Namely, the directions of their velocities are opposite. In line with the mentioned analogy, we propose the hypothesis that inside the equilibrium BH there exist low-energy excitations moving toward the exterior of the BH. These excitations are able to escape from the BH, unlike ordinary matter that falls to its center. The important consequences to the quantum theory of black holes follow.
\end{abstract}

\maketitle

\section{Introduction}

In the Painleve--Gullstrand (PG) coordinate frame, the Schwarzschild black hole solution \cite{Schwarzschild,Schwarzschild_} has a form such that the description of matter resembles the description of liquid \mbox{motion \cite{Gullstrand,Painleve}}. Such~similar reference frames also apply to the charged (Reissner--Nordstrom) and to the rotated (Kerr) black holes  \cite{Hamilton:2004au,Doran:1999gb}.~The given analogy is realized by the real motion of a superfluid~\cite{Volovik:1999fc}. In~this framework, the semiclassical calculation of Hawking radiation \cite{Hawking:1974sw} was proposed for the first time.~This calculation was revisited in \cite{Parikh:1999mf} and discussed later in a number of papers (see,~for~example,~\cite{Akhmedov:2006pg,Jannes:2011qp,Volovik2003}).~The interesting feature of the description of the black hole (BH) in the PG reference frame is that states with vanishing energy form a surface in momentum space. Therefore, in \cite{VolovikBH}, it was hypothesized that in the equilibrium state in the interior of a black hole, a Fermi surface should appear.

Recently, new materials called Weyl and Dirac semimetals were discovered \cite{semimetal_effects6,semimetal_effects10,semimetal_effects11,semimetal_effects12,semimetal_effects12_,semimetal_effects13,Zyuzin:2012tv,tewary,16}.~The dynamics of the low-energy excitations in these materials is similar to that of the elementary particles in high-energy physics.

\scalebox{.945}[1.0]{In the so-called type II Weyl {semimetals (WSII)} \cite{W2},  the dispersion of fermionic quasiparticles \cite{VZ}} has a form similar to the dispersion of the massless fermions inside a BH in the PG reference frame. This analogy has been discussed in \cite{VolovikBHW2,NissinenVolovik2017a}.~According to this scenario, a vacuum is considered as a substance similar to the superfluid component of $^3$He.~In the first stage, it falls toward the center of the BH. The excitations above this substance (i.e., elementary particles) are falling as well. Inside the horizon, the velocity of the excitations cannot be directed toward the exterior of the BH. Therefore, classical particles cannot escape from the BH. Quantum tunneling, however, leads to Hawking radiation with a thermal spectrum. The falling ``vacuum'' does not represent an equilibrium. This is why thermalization leads to the appearance of the true equilibrium.  Then,~in the Painleve--Gullstrand reference frame, fermions behave similarly to the quasiparticles a short distance from the type II Weyl point in the WSII \cite{VolovikBHW2}. There the Fermi surface is formed, which is composed of two Fermi pockets. These fermions that reside close to the type II Weyl point move in a definitive direction and cannot move in the opposite direction. However, for each type II Weyl point, a second Weyl point appears, and the quasiparticles nearby move to the direction opposite of that of the first Weyl point.

 {In \cite{Z2018}}, it was proposed that a similar pattern also occurs inside equilibrium black holes. Namely, the low-energy excitations near one piece of the Fermi surface move only toward the center of the BH, while the low-energy excitations near the other piece of the Fermi surface move only toward the exterior of the black hole.~This hypothesis has been illustrated by a particular lattice model, which belongs to a class of the analogue gravity models (for discussion of the analogue gravity models, see, for example, {\cite{Ge:2015uaa}} {and references therein}).

It is worth mentioning, however, that contrary to the models like that of \cite{Ge:2015uaa}, we do not discuss the dynamics of the gravitational field itself. We only deal with the dynamics of matter in the presence of a fixed gravitational background and completely neglect the back reaction. This allows us to seriously take into account the analogy with type II Weyl semimetals while considering the motion of elementary particles. Below, we review the approach of \cite{Z2018} without specifying the particular toy model. We rely only on the compactness and finiteness of the Fermi surface formed inside the black hole according to the scenario of \cite{VolovikBH}.

\section{Dirac Fermions in the BH in PG Reference Frame}
\label{Section_DF}

The Reissner--Nordstrom BH metric in these coordinates is of {the form}
\begin{equation}
ds^2 = dt^2 - (d{\bf r} - {\bf v}({\bf r}) dt)^2, \label{PGQ}
\end{equation}
where
\begin{equation}
{\bf v} = -\frac{1}{m_P} \frac{\bf r}{r}\, \sqrt{\frac{2M}{r}-\frac{Q^2}{r^2}}
\end{equation}
and, in a certain sense, may be considered as the velocity of a falling substance. We denote by $Q$ the BH charge, by $m_p$ the Plank mass, and by $M$ the BH mass.

Vielbein is in the form
\begin{equation}
E^\mu_a = \left(\begin{array}{cc}1 & {\bf v}\\0 & 1\end{array} \right)
\end{equation}
the inverse matrix to $E^\mu_a$ is
\begin{equation}
e_\mu^a = \left(\begin{array}{cc}1 & -{\bf v}\\0 & 1\end{array} \right)
\end{equation}
and the metric is defined as
$$ g_{\mu \nu} = e^a_\mu e^b_\nu \eta_{ab}$$
where $\eta_{ab} = {\rm diag}\,(1,-1,-1,-1)$ is the metric of Minkowski space.

Massless Weyl fermion {action is given by}
\begin{eqnarray}
S_{R,L} &=& \int d^4x \, {\rm det}^{-1}({\bf E})\,\bar{\psi}_{R,L}(x) \Big(i  E^0_0\partial_t \pm i E^k_a \tau^a \partial_k  \Big) \psi_{R,L}(x) \nonumber\\ &=& \int d^4x \, \bar{\psi}_{R,L}(x) \Big(i \partial_t - H^{R,L}(-i \partial) \Big) \psi_{R,L}(x)
\end{eqnarray}
where the subscript $R$ ($L$) marks the right-handed/left-handed fermions, while
\begin{equation}
H^{R,L}({\bf p}) = \pm {\bf p} \sigma {+} {\bf p} {\bf v}
\end{equation}

Adding the mass term and mixing the right-handed and the left-handed fermions, \begin{eqnarray}
S_m &=& - m \sum_{R,L}\int d^4x \, {\rm det}^{-1}({\bf E})\,\bar{\psi}_{R,L}(x) \psi_{L,R}(x)\nonumber\\ &=& - m \int d^4x \,\sum_{R,L} \bar{\psi}_{R,L}(x)  \psi_{L,R}(x)
\end{eqnarray}

we arrive at the description of Dirac fermions. It is worth mentioning that the spin connection does not manifest itself in this action in the particular case of the considered BH background.

The two horizons are placed at
\begin{equation}
r_+=\frac{M + \sqrt{M^2 - Q^2 m_P^2}}{m_P^2}
\end{equation}
and
\begin{equation}
r_-=\frac{M - \sqrt{M^2 - Q^2 m_P^2}}{m_P^2}
\end{equation}

For $ r > r_+$, there are ordinary Dirac fermions. At $m=0$, the Fermi point appears. Between the two horizons, $r_- < r < r_+$ at $m=0$, there is the type II Dirac point, while $|{\bf v}|$ is larger than light velocity.

For $r < r_0 = \frac{Q^2}{2M}$, velocity $v$ becomes imaginary. We consider this unphysical and suppose that in the final theory, due to interaction with matter, the dependence of $v$ on $r$ is modified within the BH so that $v$ remains real and vanishes at $r=0$ only. More details about the possible modification of Reissner--Nordstrom geometry at small $r$ may be found in \cite{Hamilton:2004au,Hamilton:2008zz}.

\section{What Is Happening If the Equilibrium Is Achieved?}
\label{Secteq}

In the conventional state of the BH, the ``vacuum'' may be considered as a substance falling to the center of the BH. The simplification accepted here is that this substance looks as though the particles do not interact with each other. Each particle has velocity ${\bf v}({\bf r})$ (measured in self-time). Locally, in the falling reference frame, space--time is flat, and the Dirac particles with excitations above that of the vacuum have the dispersion $$ E(p) = \pm \sqrt{p^2 + m^2.}$$

The lower branch of the spectrum is occupied.

The next step is the consideration of the vacuum as a substance with nontrivial properties. Such an attempt was made with the q-theory {proposed in}
 \cite{Klinkhamer:2016jrt}. However, here we do not rely on this theory and proceed to consider the falling vacuum as a collection of occupied states corresponding to noninteracting particles. We suppose that, qualitatively, this description is enough for our purposes.

\begin{figure}
\begin{center}
{\epsfig{figure=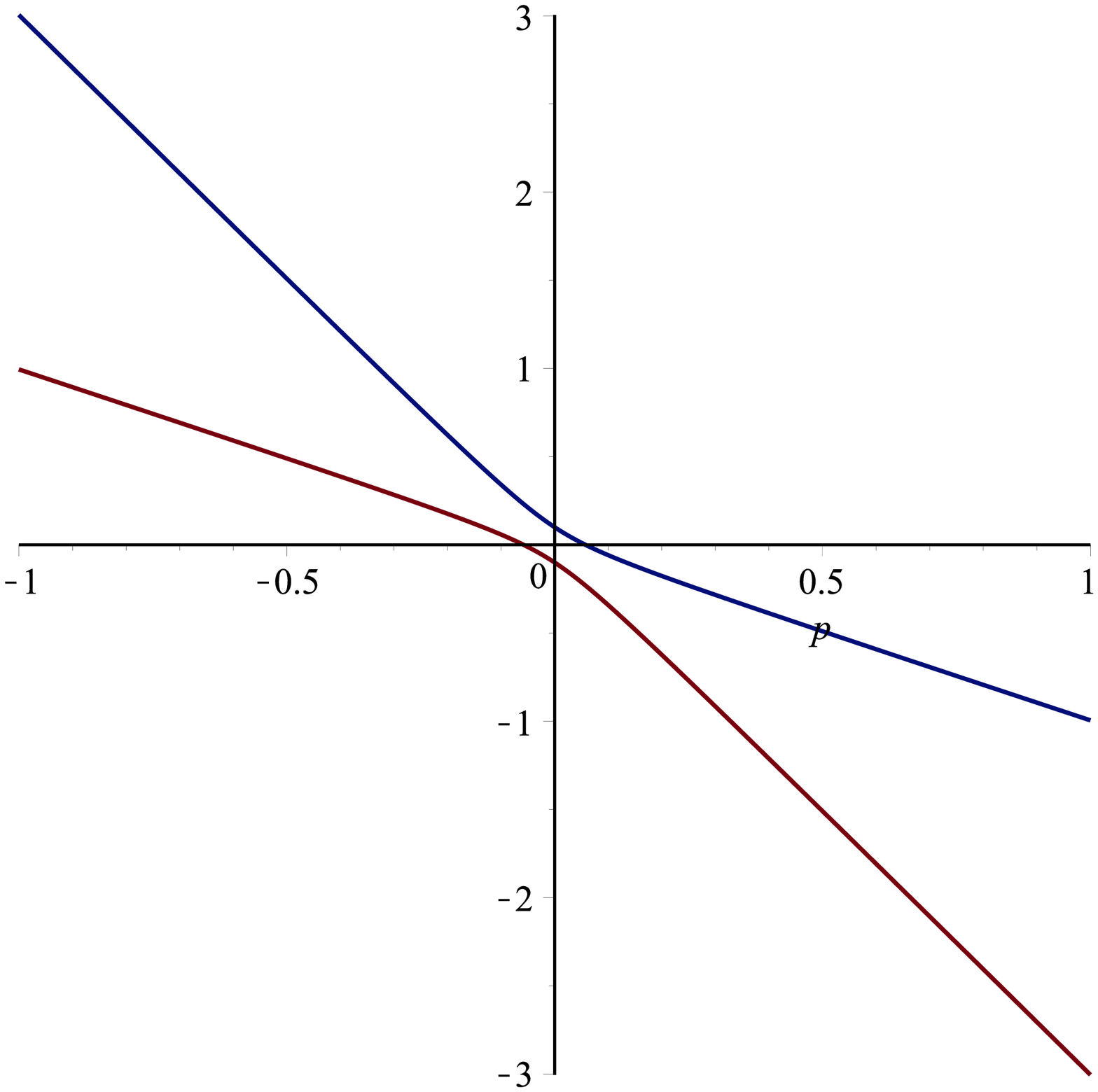,width=35mm,angle=0}}
{\epsfig{figure=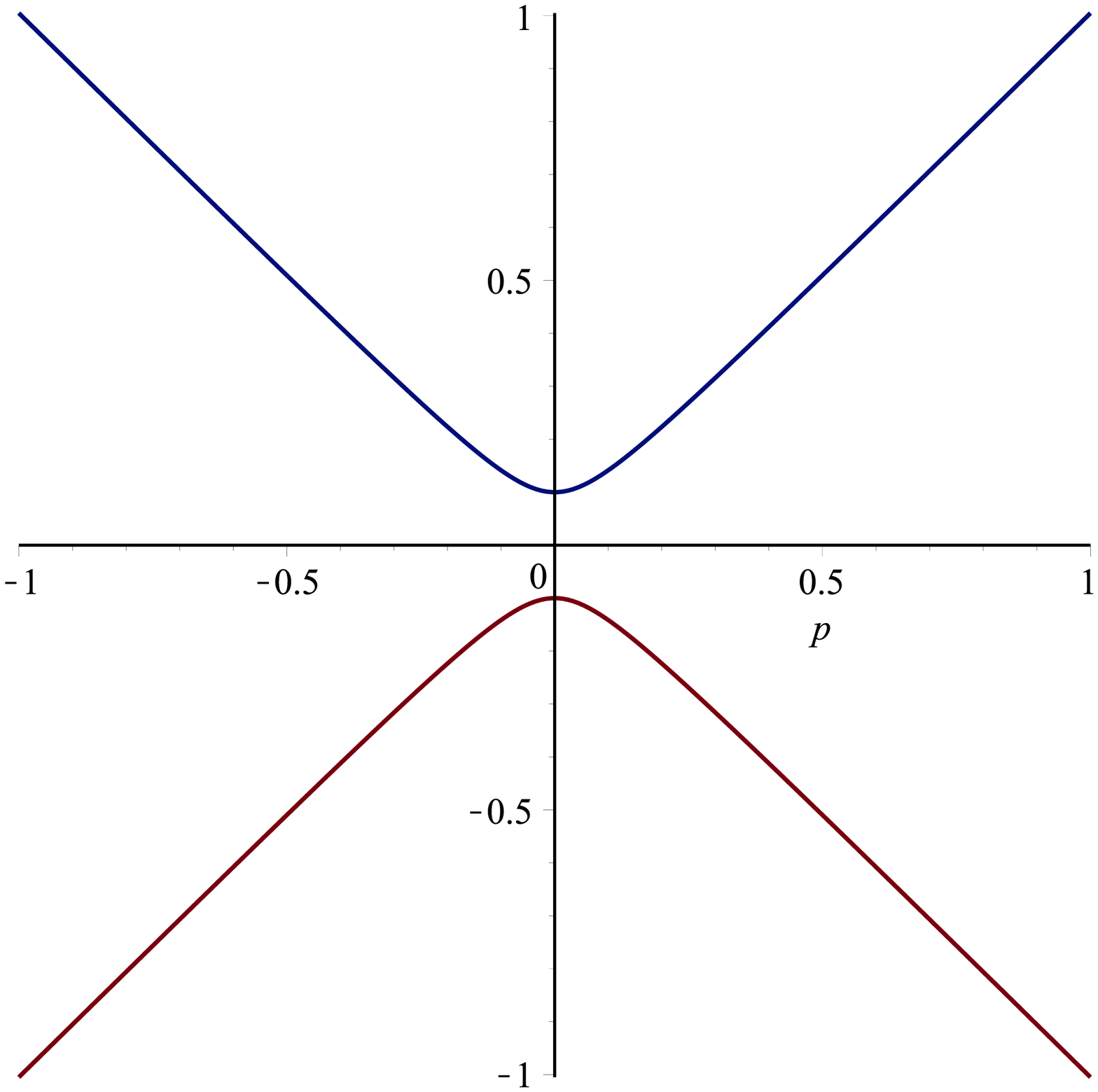,width=35mm,angle=0}}
\end{center}
\caption{Energy of Dirac fermions as a function of momentum in the Painleve--Gullstrand (PG) coordinate frame. The vertical axis corresponds to energy while the horizontal axis corresponds to momentum.~({\bf Left})  Inside the horizon of the BH. The states along the lower line are occupied. There is a piece of this line above zero. The~corresponding particles tunnel through the horizon toward the exterior of the BH. ({\bf Right}) Outside of the external horizon. Again, the lower branch is occupied, but it remains completely below~zero.}
\label{fig.0}
\end{figure}

In the PG coordinate frame, the local dispersion of fermionic particles is
\begin{equation} E({\bf p}) = \pm \sqrt{p^2 + m^2} {+} {\bf v} {\bf p}\label{EE}\end{equation}
At $v>1$, the Dirac cone is overtilted. Therefore, there is part of the occupied branch of the spectrum with energies above zero (see Figure \ref{fig.0}). This is why the fermions are able to tunnel from the interior of the BH to its exterior. This is called Hawking radiation  \cite{Volovik:1999fc,VolovikBH,Akhmedov:2006pg,Jannes:2011qp}.

Locally, the above-mentioned substance falling toward the center of the BH looks like a vacuum in the coordinate system falling toward the BH center.~Globally, this state is not in equilibrium. Next, we must take interactions into account. They drive the system to the true equilibrium. After thermalization, the system comes to a state with finite temperature that is composed of the occupied states with energies below zero (defined in the PG reference frame) and thermal quasiparticles.

\begin{figure}
\centering
 \epsfig{figure=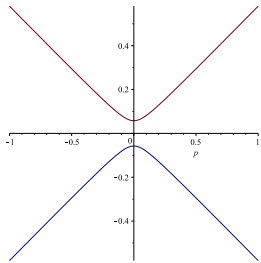,width=35mm,angle=0}
\caption{Fermi pockets are presented for the model with Dirac fermions, with $m=0.1$ and $v=2$. The vertical axis corresponds to radial momentum while the horizontal axis corresponds to the transversal component of momentum.}
\label{fig.1}
\end{figure}

At an infinite distance from the BH {there} are ordinary Dirac fermions. When we approach the BH, the corresponding Dirac cone is tilted, but there is still a mass gap. At the horizon, the gap disappears. Inside the BH
(for $r_- < r < r_+$), the dependence of energy on momentum is given by Equation (\ref{EE})
with $|{\bf v}|>1$. There, the Dirac cone is overtilted. Thus, as mentioned above, in the original nonequilibrium state, there is a piece of the branch of the spectrum above zero that is filled with particles, while there also appears a piece of the branch below zero with vacant states. After thermalization, vacuum reconstruction occurs: the states below zero are all occupied except for the thermally excited holes, and the states above zero are all vacant except for the thermally excited particles. As a result, the two pieces of the Fermi surface are formed, as represented in {Figure \ref{fig.1}}.~This pattern is similar to the description of the vicinity of each type II Weyl point in a type II Weyl semimetal \cite{W2}. The analogy becomes complete for the case of vanishing mass when the two pieces of the Fermi surface touch each other at the Weyl point. These two pieces are called Fermi pockets. In the presence of low mass, a small distance appears between the Fermi pockets.

The Standard Model (SM) fermions reside within a small area of the point $P^{(0)} = 0$. Outside of the BH, the values of energies for momenta sufficiently distant from this point are large. The corresponding states cannot be excited at low energies. However, inside the BH, the majority of the Fermi surface is situated far from  $P^{(0)}$. The corresponding states are all relevant at low energies. We come to the conclusion that the interior of the BH is to be described by the ultraviolet completion of the SM, even at low energies.

The particles of the Standard Model fall down to the BH. They pass through the two horizons and enter the region $r<r_-$.~There, according to the pattern of the analytically continued Reissner--Nordstrom geometry, they may enter a new Universe.~We, however, assume that this pattern is to be modified and there is no entrance to any new World inside the BH. What will then happen to the SM particles that have fallen to the BH?
\section{Excitations That Are Able to Escape from the BH Directly, Not through the Tunneling Mechanism}

A guess at the last question in the previous section is that the SM particles in the interior of the BH may be transformed to the excitations of a more general theory than the SM. The next guess is that the latter are able to move toward the exterior of the outer horizon.

\begin{figure}
\centering
\includegraphics[width=4.5cm]{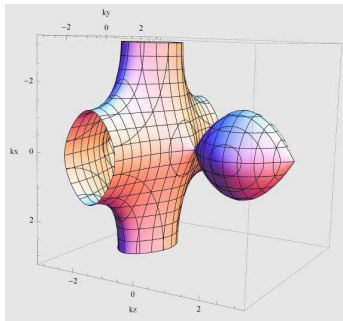}
\caption{The typical form of the Fermi surface for a type II Weyl
semimetal described by the toy model of \cite{ZL2018}. The two Fermi pockets touch each other at the two type II Weyl points. One Fermi pocket has the topology of a sphere while another has the topology of a Riemann surface with two handles. The three axes of the figure correspond to the three components of momenta.}
\label{fig3}
\end{figure}

\begin{figure}
\centering
\includegraphics[width=4.5cm]{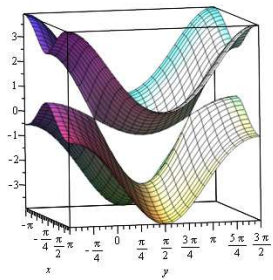}
\caption{Dispersion of the quasiparticles (energy as a function of momenta $k_x=k_1, k_z=k_3$) in the toy
model of \cite{ZL2018}. The vertical axis corresponds to energy, the two horizontal axes correspond to the two components of momenta.}
\label{fig.4}
\end{figure}

Bearing in mind the analogy to type II Weyl semimetals, let us consider, in more detail, what is happening there in equilibrium. In \cite{ZL2018}, {a toy} 
model was proposed that reflects the characteristic features of certain real WSII. The Fermi surface in this toy model is presented in Figure \ref{fig3}. It consists of the two pieces that touch each other at the two Fermi points. A similar pattern takes place for the type II Weyl semimetal WP$_2$ (Figure 1b of \cite{WP2}): the two pairs of the type II Weyl points  are connected by two Fermi pockets. Fermionic excitations that reside in the vicinities of the opposite type II Weyl points move in opposite directions because  the gradients of energy (as a function of momenta) there are directed oppositely (this is illustrated by Figure \ref{fig.4}).

\unskip

In general, a similar situation takes place for other real type II Weyl semimetals. In particular, in $\gamma - $MoTe$_2$ (see Figure 3 of \cite{gMoTe2}), the pair of type II Weyl points belongs to the common piece of the Fermi surface. However, there are also extra Fermi pockets that are not common to those type II Weyl points. Moreover, there are separate pieces of the Fermi surface that are distant from the Weyl points.
For each piece of the Fermi surface, there are parts where the gradients of the dependence of energy on momenta are directed oppositely. Thus, there are low-energy quasiparticles with opposite directions of velocity.

The central assumption of the present paper about the physics of the interior of a BH is that its description is similar to that of real type II Weyl semimetals. This is why we suppose that the Fermi surface that appears in the equilibrium state within the BH interior is closed. Then, there are pieces of the Fermi surface where the velocity of the quasiparticles is directed toward the exterior of the BH.

\begin{figure}
\centering
\epsfig{figure=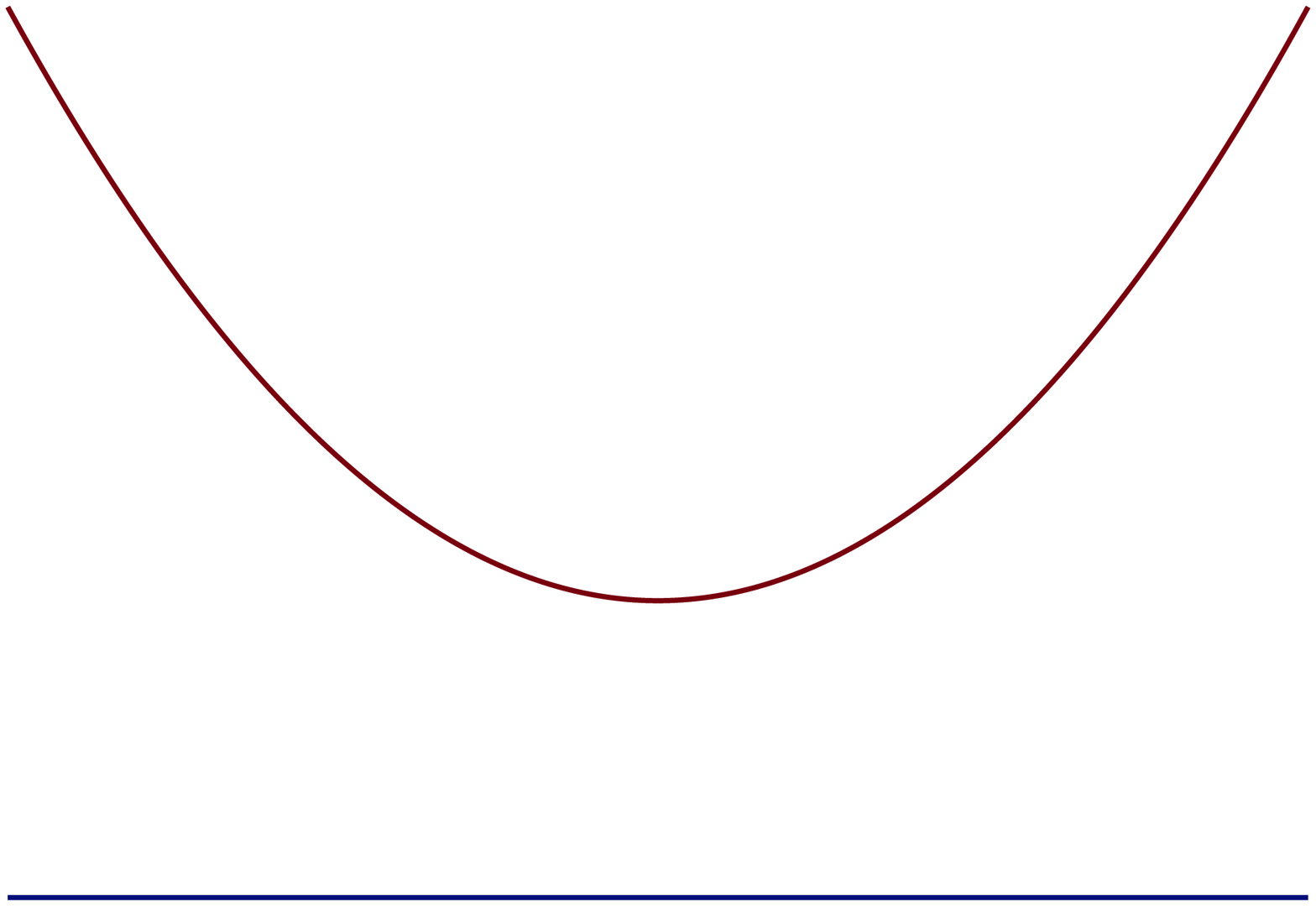,width=30mm,angle=0}
\epsfig{figure=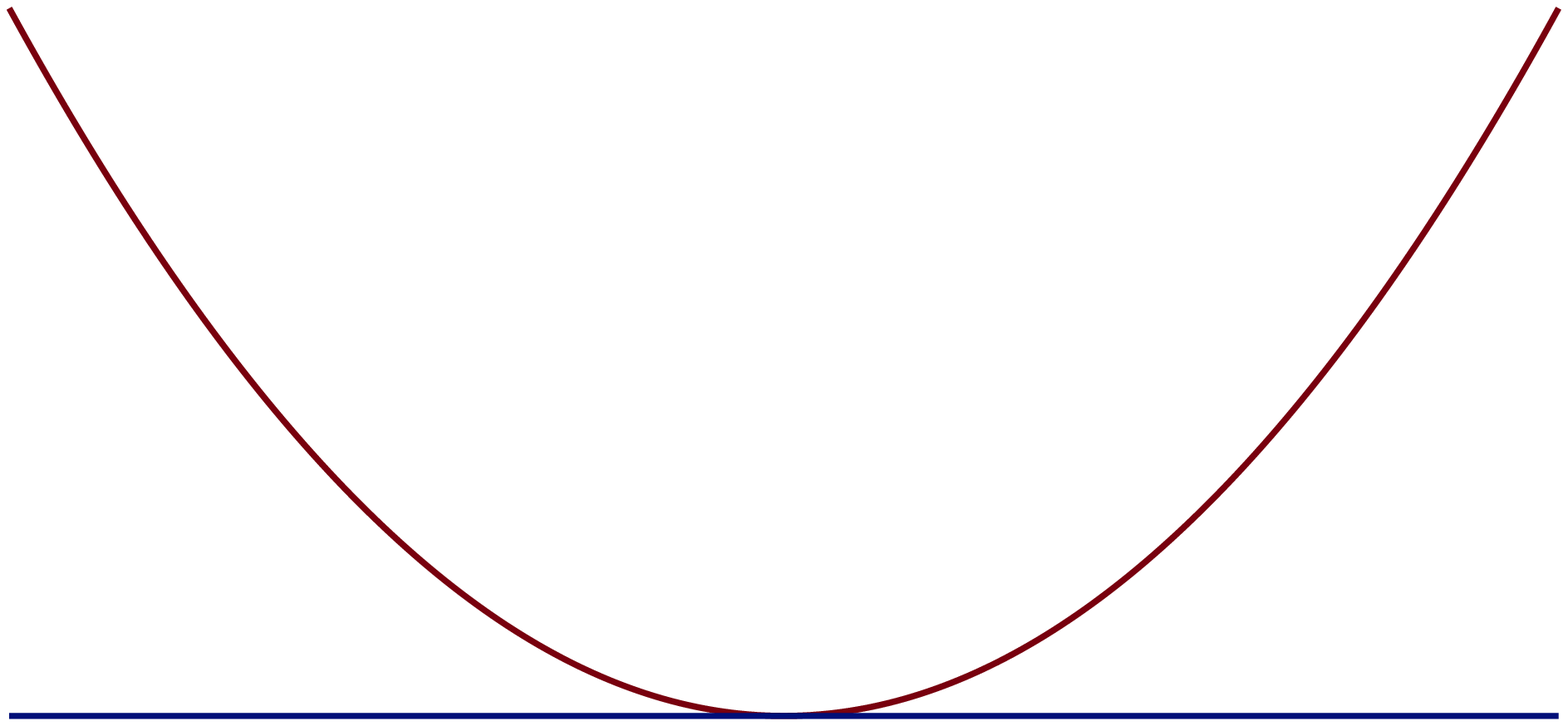,width=30mm,angle=0}
\epsfig{figure=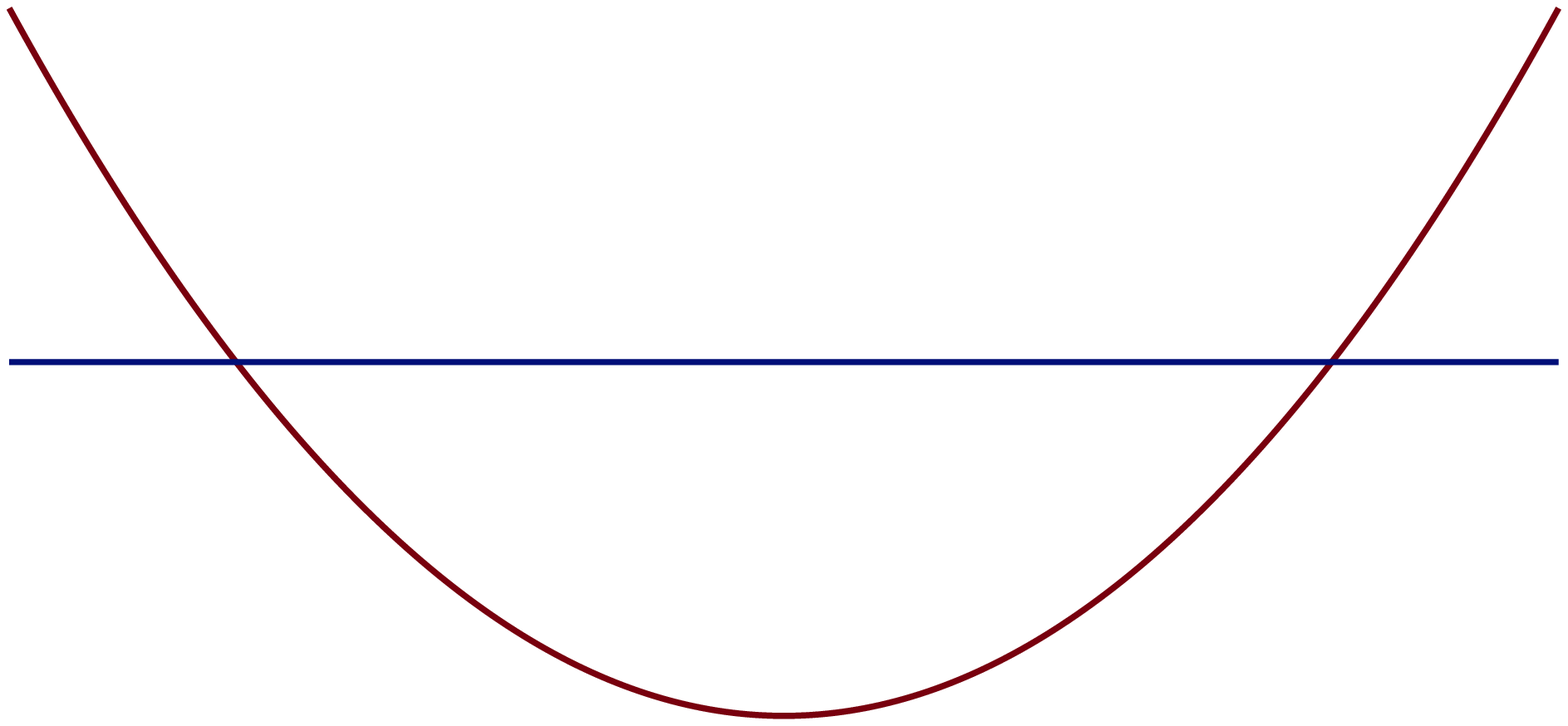,width=30mm,angle=0}
\caption{Spectrum branch for the fermionic quasiparticles.~The vertical axis is energy. The horizontal axis schematically represents the momentum.~({\bf Left}) Outside of the BH. ({\bf Center}) At the horizon. ({\bf Right}) Inside the BH.}
\label{fig.67}
\end{figure}

The sketch of the proof is as follows.~We consider the branch of the spectrum of matter (the~dependence of energy on the $z$ component of momentum) that is represented schematically in Figure \ref{fig.67}, left. This pattern corresponds to the gapped (massive) matter existing far from the BH. This branch of the spectrum may have several minima corresponding to several types of particles. Some of those minima may have extremely large values corresponding to the extra massive matter. Some may have relatively small values representing the ordinary SM matter. (This complication is illustrated by the toy model of \cite{Z2018}.) However, those details do not matter for the pattern of the Fermi surface formation presented below. We move along axis $z$ toward the center of the BH. At a certain moment, the gap disappears. This is the position of the horizon (Figure \ref{fig.67}, center). Inside the horizon, the considered branch of the spectrum crosses zero (Figure \ref{fig.67}, right). One can see that, at the crossing points, the slopes  have opposite signs. The left crossing point schematically represents the SM excitations, which move only toward the center of the BH. Correspondingly, the right crossing point represents the excitations moving in the opposite direction.

Additional matter cannot be observed far outside of the BH because of the large energy gap, which is because their masses are much larger than the masses of the SM particles. However, it may escape from the BH and be observed in the vicinity of the horizon.

\section{Conclusions}

\label{Sect_CON}

In the present paper, we rely on the description of the black hole in Einstein's theory of gravity, which enables the use of the Painleve--Gullstrand reference frame, where the metric has the especially simple form of Equation (\ref{PGQ}). Quantum fields may be considered on the same grounds in a similar reference frame  for a rotating black hole (given in \cite{Hamilton:2004au}). Besides, it would be interesting to consider a similar reference frame and the corresponding dynamics of fields for the black hole solutions in the massive gravity models (see, for example, \cite{Arraut:2014uza,Arraut:2014sja,Arraut:2015dva} and references therein). However, this remains beyond scope of the present paper.

To conclude, the circle of the life of a BH is described as follows.~At the beginning of its formation, the BH appears in a non-equilibrium state. This is the conventional state of a BH accompanied by Hawking radiation. Next, according to the scenario in \cite{VolovikBH}, the vacuum of the black hole is reconstructed: the black hole arrives at the equilibrium state with essentially different properties. As mentioned above, we do not know how much time that vacuum reconstruction requires. Obviously, for {a} particular BH, this thermalization time should be {compared} to the time of its existence as well as to the evaporation time. If it is much larger than both these timescales, then this particular BH never realizes the equilibrium state. Suppose, however, that this transition does occur.

Then, we explore the analogy between the description of electronic quasiparticles in type~II Weyl semimetals and the description of Dirac fermions inside the BH in the PG reference frame. According to this analogy, the SM quasiparticles reside in the vicinity of the point $P^{(0)}$ in momentum space. In equilibrium, inside the BH, the Fermi surface is formed, and there are pieces of this Fermi surface that are situated far from point $P^{(0)}$. The slope of the energy dispersion close to $P^{(0)}$ inside the BH is such that the quasiparticle velocity is directed toward the center of the BH. This is why ordinary SM matter cannot escape from the BH without tunneling. On the other hand, there exist parts of the Fermi surface where the quasiparticle velocity has the opposite direction. The corresponding quasiparticles inside the BH have small energies and move toward the exterior of the BH. Such low-energy excitations may escape and exist outside of the horizon{.}

Strictly speaking, the hypothesis about the interior of the BH proposed above poses more questions than it answers. In particular, in order to take into account the particles that escape directly from the BH, we are likely to essentially change BH thermodynamics, which is essentially based on the assumption that nothing escapes from a BH except for Hawking radiation. It is not clear to what extent the conclusions {of} BH thermodynamics will survive and to what extent they are to be corrected. This is to be the subject of a separate study. It is worth mentioning that the pattern described here implies that only SM matter interacts with gravity in a conventional way. The~excitations that escape from the BH obviously interact with the gravitational field in a different way. This is why it is also not reasonable to assume that the usual Einstein equations with contributions from all available excitations determine the gravitational field completely (at least, inside the BH and in its vicinity).

At the same time, we hope that the above-presented scheme may be able to {resolve} the so-called information paradox \cite{information,BHI}: the information does not disappear in the black hole after its evaporation if, on the way, the equilibrium state of the black hole is approached, and the mentioned extra massive matter escapes from inside the horizon carrying the would-be missed information. This~conclusion appears very naturally because the BH information paradox appears within the ranges of the non-regularized quantum field theory (QFT) defined for the BH background. Our consideration is based on the analogy between the QFT and the electronic structure of solids. The other formulation of this analogy is the consideration of the QFT on the lattice. Thus, the information paradox is likely to disappear when the QFT is regularized properly.


\vspace{6pt}





\acknowledgments{The author kindly acknowledges useful discussions with {Volovik, G.E.} } 







\begin{thebibliography}{99}


\bibitem{Schwarzschild}  {Schwarzschild, K. Uber das Gravitationsfeld eines Massenpunktes nach der Einsteinschen Theorie (1916)}.   \emph{arXiv} \textbf{1999}, arXiv preprint physics/9912033.  


\bibitem{Schwarzschild_}
Schwarzschild, K. {Uber das Gravitationsfeld einer Kugel aus inkompressibler Flussigkeit nach der Einsteinschen Theorie. Available online: http://adsabs.harvard.edu/abs/1916skpa.conf..424S (accessed on 05 May 2018).} 




\bibitem{Gullstrand}
Gullstrand, A. Allgemeine Losung des statischen Einkorperproblems in der Einsteinschen Gravitationstheorie. \emph{Arkiv. Mat. Astron. Fys.} \textbf{1922}, \emph{16}, 1--15.

\bibitem{Painleve}
Painleve, P. La mecanique classique et la theorie de la relativite. \emph{C. R. Acad. Sci. } \textbf{1921}, \emph{173}, 677--680.

\bibitem{Hamilton:2004au}
Hamilton,   A.J.S.; Lisle, J.P. The River model of black holes.  \emph{Am. J. Phys.} \textbf{2008}, {\em 76}, 519--532.



\bibitem{Doran:1999gb}
Doran, C. A New form of the Kerr solution.  \emph{Phys. Rev. D} \textbf{2000}, {\em 61},  067503.

\bibitem{Volovik:1999fc}
Volovik,   G.E. Simulation of Painleve-Gullstrand black hole in thin He-3---A film. \emph{JETP Lett.} \textbf{1999}, {\em 69}, 705,  doi:10.1134/1.568079.

\bibitem{Hawking:1974sw}
Hawking, S.W. Particle Creation by Black Holes. \emph{Commun. Math. Phys. } \textbf{1975}, {\em 43}, 199--220.


\bibitem{Parikh:1999mf}
Parikh,   M.K.; Wilczek, F. Hawking radiation as tunneling.  \emph{Phys. Rev. Lett. } \textbf{2000}, {\em 85}, 5042--5045.

\bibitem{Akhmedov:2006pg}
Akhmedov,   E.T.; Akhmedova, V.; Singleton, D. Hawking temperature in the tunneling picture.  \emph{Phys. Lett. B} \textbf{2006}, {\em 642}, 124--128.

\bibitem{Jannes:2011qp}
  Jannes, G. Hawking radiation of $E < m$ massive particles in the tunneling formalism.  \emph{JETP Lett.}  \textbf{2011}, {\em 94}, 18, doi:10.1134/S0021364011130091.

\bibitem{Volovik2003}
Volovik, G.E. {\it The Universe in a Helium Droplet}; Clarendon Press:  Oxford, UK, 2003.

\bibitem{VolovikBH}
  Huhtala, P.; Volovik, G.E. Fermionic microstates within Painleve-Gullstrand black hole. \emph{J. Exp. Theor. Phys.}   \textbf{2002}, {\em 94}, 853, doi:10.1134/1.1484981.


\bibitem{semimetal_effects6}
Parameswaran, S.; Grover, T.; Abanin, D.; Pesin, D.; Vishwanath,  A. Probing the chiral anomaly with nonlocal transport in Weyl semimetals. \emph{Phys. Rev. X} \textbf{2014}, {\em 4}, 031035.


\bibitem{semimetal_effects10}
Vazifeh, M.; Franz, M. Electromagnetic response of weyl semimetals. \emph{Phys. Rev. Lett.} \textbf{2013}, {\em 111}, 027201.

\bibitem{semimetal_effects11}
Chen, Y.; Wu, S.;  Burkov, A. Axion response in Weyl semimetals. \emph{Phys. Rev. B} \textbf{2013}, {\em 88}, 125105.

\bibitem{semimetal_effects12}
Chen, Y.; Bergman, D.;  Burkov, A. Weyl fermions and the anomalous Hall effect in metallic ferromagnets.
\emph{Phys. Rev. B} \textbf{2013}, {\em 88}, 125110

\bibitem{semimetal_effects12_}
{Vanderbilt, D.; Souza, I.; Haldane, F.D.M. } Comment on “Weyl fermions and the anomalous Hall effect in metallic ferromagnets”. \emph{Phys. Rev. B} \textbf{2014}, {\em 89}, 117101.


\bibitem{semimetal_effects13}
Ramamurthy, S.T.; Hughes, T.L. Patterns of electro-magnetic response in topological semi-metals.  \emph{Phys. Rev. B} \textbf{2015}, \emph{92}, 085105.


\bibitem{Zyuzin:2012tv}
Zyuzin, A.A.; Burkov, A.A. Topological response in Weyl semimetals and the chiral anomaly.  \emph{Phys. Rev. B} \textbf{2012}, {\em 86}, 115133.


\bibitem{tewary}
Goswami, P.; Tewari, S. Axionic field theory of (3 + 1)-dimensional Weyl semi-metals. \emph{Phys. Rev. B} \textbf{2013}, \emph{88}, 245107.

\bibitem{16}
Liu, C.-X.; Ye,  P.; Qi, X.-L. Chiral gauge field and axial anomaly in a Weyl semimetal. \emph{Phys. Rev. B} \textbf{2013}, \emph{87}, 235306.


\bibitem{W2}
 Soluyanov, A.A; Gresch, D.; Wang, Z.; Wu, Q.S.; Troyer, M.; Dai, X.; Bernevig, B.A. Type-II Weyl Semimetals. \emph{Nature} \textbf{2015}, \emph{527}, 495--498.


\bibitem{VZ}
Volovik, G.E.;  Zubkov, M.A. Emergent Weyl spinors in multi-fermion systems. \emph{Nucl. Phys. B} \textbf{2014}, \emph{881}, 514--538.


\bibitem{VolovikBHW2}
Volovik,  G.E. Black hole and Hawking radiation by type-II Weyl fermions. \emph{JETP Lett.}  \textbf{2016}, {\em 104},  645--648.



\bibitem{NissinenVolovik2017a}
Nissinen, J.;  Volovik, G.E. Type-III and IV interacting Weyl points. \emph{JETP Lett.} \textbf{2017}, {\em 105},  447--452.

\bibitem{Z2018}
Zubkov, M.A. The black hole interior and the type II Weyl fermions.  \emph{Mod. Phys. Lett. A} \textbf{2018}, {\em 33},  1850047.


\bibitem{Ge:2015uaa}
Ge,  X.H.; Sun, J.R.; Tian, Y.; Wu, X.N.; Zhang, Y.L. Holographic Interpretation of Acoustic Black Holes.  \emph{Phys.~Rev. D} \textbf{2015}, {\bf 92},  084052.


\bibitem{Hamilton:2008zz}
Hamilton,   A.J.S.; Avelino, P.P. The Physics of the relativistic counter-streaming instability that drives mass inflation inside black holes. \emph{ Phys. Rep.} \textbf{2010}, {\em 495}, 1--32.


\bibitem{Klinkhamer:2016jrt}
Klinkhamer, F.R.; Volovik, G.E. Propagating q-field and q-ball solution. \emph{Mod. Phys. Lett. A}  \textbf{2017}, {\em 32},  1750103.



\bibitem{ZL2018}  Zubkov, M.A.; Lewkowicz, M. The type II Weyl semimetals at low temperatures: Chiral anomaly, elastic deformations, zero sound. \emph{Ann. Phys.} \textbf{2018}, \emph{399}, 26--52.


\bibitem{WP2}
Schonemann, R.; Aryal, N.; Zhou, Q.;  Chiu, Y.-C.; Chen, K.-W.; Martin, T.J.; McCandless, G.T.;  Chan, J.Y.;  Manousakis, E.;  Balicas, L. Fermi surface of the Weyl type-II metallic candidate WP$_2$. \emph{Phys. Rev. B} \textbf{2017}, \emph{96},~121108.

\bibitem{gMoTe2}
Rhodes, D.; Schonemann, R.; Aryal, N.;  Zhou, Q.; Zhang,  Q.R.; Kampert, E.; Chiu, Y.-C.; Lai, Y.; Shimura, Y.;  McCandless,  G.T.; et al. Bulk Fermi-surface of the Weyl type-II semi-metallic candidate MoTe2. \emph{Phys. Rev. B} \textbf{2017}, \emph{96}, 165134.





\bibitem{information} Page,  D.N. Black hole information. \emph{arXiv} \textbf{1995}, arXiv:hep-th/9305040. Available online: https://arxiv.org/abs/hep-th/9305040 (accessed on  {5 November 2018}).

\bibitem{BHI} Harlow, D. Jerusalem Lectures on Black Holes and Quantum Information. \emph{Rev. Mod. Phys.} \textbf{2016}, \emph{88}, 15002.



\bibitem{Arraut:2014uza}
Arraut,  I. The Black Hole Radiation in Massive Gravity.  \emph{Universe} \textbf{2018}, {\em 4},  27.


\bibitem{Arraut:2014sja}
Arraut,   I. On the apparent loss of predictability inside the de-Rham-Gabadadze-Tolley non-linear formulation of massive gravity: The Hawking radiation effect.  \emph{EPL (Europhys. Lett.)} \textbf{2015}, {\em 109},  10002.


\bibitem{Arraut:2015dva}
 Arraut,  I. Path-integral derivation of black-hole radiance inside the de-Rham-Gabadadze-Tolley formulation of massive gravity.  \emph{ Eur. Phys.  J. C} \textbf{2017}, {\em 77},  501.



\end{thebibliography}
\end{document}